\documentclass{article}
\usepackage{graphicx} 
\usepackage{hyperref}%
\usepackage{cite}
\usepackage{algorithm}
\usepackage{algpseudocode}
\usepackage{amsmath, algorithm, algpseudocode}
\usepackage{graphicx} 
\usepackage{graphicx} 
\usepackage{amssymb}
\usepackage{amsmath}
\usepackage{amsthm}
\usepackage{color}
\usepackage{comment}
\usepackage{enumitem} 
\usepackage{geometry} 
\usepackage{authblk}
\usepackage{amssymb}
\usepackage{listings}
\usepackage{xcolor}
\usepackage{caption}
\usepackage{color}
\usepackage{algorithm}
\usepackage{algpseudocode}
\usepackage{algorithmicx}
\usepackage{subcaption}

\theoremstyle{definition}

\theoremstyle{remark}

\title{DICOM Compatible, 3D Multimodality Image Encryption using Hyperchaotic Signal}
\author{
    Anandik N Anand\textsuperscript{1}, 
    Sishu Shankar Muni\textsuperscript{1}\thanks{*Corresponding author: sishushankarmuni@gmail.com}, 
    Abhishek Kaushik\textsuperscript{2} \\
    \small \textsuperscript{1}School of Digital Sciences, Digital University Kerala,\\
    \small Thiruvananthapuram, Kerala, India \\
    \small \textsuperscript{2}School of Computer Science and Engineering, Digital University Kerala,\\
    \small Thiruvananthapuram, Kerala, India
}

\date{}

\begin{document}

\maketitle
\begin{abstract}

Medical image encryption plays an important role in protecting sensitive health information from cyberattacks and unauthorized access. In this paper, we introduce a secure and robust encryption scheme that is multi-modality compatible and works with MRI, CT, X-Ray and Ultrasound images for different anatomical region of interest. The method utilizes hyperchaotic signals and multi-level diffusion methods. The encryption starts by taking DICOM image as input, then padding to increase the image area. Chaotic signals are produced by a logistic map and are used to carry out pixel random permutation. Then, multi-level diffusion is carried out by 4-bit, 8-bit, radial and adjacent diffusion to provide high randomness and immunity against statistical attacks. In addition, we propose a captcha-based authentication scheme to further improve security. An algorithm generates  alphanumeric captcha-based image which is encrypted with the same chaotic and diffusion methods as the medical image. Both encrypted images(DICOM image and captcha image) are then superimposed to create a final encrypted output, essentially integrating dual-layer security. Upon decryption, the superimposed image is again decomposed back to original medical and captcha images, and inverse operations are performed to obtain the original unencrypted data. Experimental results show that the proposed method provides strong protection with no loss in image integrity, thereby reducing unauthorized data breaches to a significant level. The dual-encryption approach not only protects the confidentiality of the medical images but also enhances authentication by incorporating captcha.

\end{abstract}
\section*{Introduction}

Medical images form a vital part of contemporary medicine, and their role in the diagnosis, planning of treatment, and follow-up of medical disorders cannot be underestimated. MRI, CT, X-ray, PET, and ultrasound are some of the imaging modalities focused in this work.
With the ever-accelerating development of healthcare technologies and the growing deployment of digital imaging, the amount of medical data produced has risen exponentially. Since medical images do carry extremely sensitive patient information, their confidentiality and integrity become a major issue to secure. Data breaches in medical information can have serious implications, such as identity theft, loss of patient confidence, and privacy regulation violations like the Health Insurance Portability and Accountability Act (HIPAA)\cite{act1996health} and the General Data Protection Regulation (GDPR)\cite{regulation2018general}.

Encryption is a basic method of protecting digital data by rendering it unreadable, which can be decrypted by only authorized individuals who have the proper key. In medical imaging, encryption plays a critical role in ensuring sensitive information is secured during transmission and storage. Standard encryption techniques like the Advanced Encryption Standard (AES)\cite{nechvatal2001report} and Data Encryption Standard (DES)\cite{standard1999data}are widely applied to general data protection. However, these methods are not fully appropriate for medical images because they are of high dimension and possess specific characteristics, such as spatial correlation and redundancy between neighboring pixels.

Medical images, in general, being large and complex in structure, are challenging in terms of securing high encryption rate with high resistance to attacks. Additionally, medical images tend to be highly redundant, thus highly susceptible to statistical and differential attacks if not sufficiently encrypted\cite{amin2007statistical}. Hence, there is increasing demand for techniques of encryption which not only offer security but also preserve the image quality and its diagnostic worth\cite{ding2020deepedn}.

Decryption is the opposite operation of converting encrypted data into its original, readable state. Decryption, in medical imaging, must be fast and accurate such that the reconstructed image possesses its diagnostic worth. Any deterioration in image integrity during decryption could lead to flawed diagnoses or substandard medical interpretations. The decryption process, therefore, should be fault-free and computationally efficient such that the original image quality is preserved\cite{wu2019computationally}.

Medical images are now being transmitted across public and private networks, leaving them vulnerable to unauthorized access and manipulation. With the increased use of telemedicine, healthcare information systems, and cloud storage, encryption has become a necessity to protect medical data\cite{kondakov2020intelligent},\cite{doukas2010mobile}. The implications of breached medical images are severe, as they can result in patient data exposure and even alteration of medical records. Therefore, those techniques which cover both security as well as computational efficiency are much sought after\cite{shankar2018efficient}.

Chaotic systems, owing to their sensitive dependence on initial conditions and random-like behavior, have been suggested as effective tools for secure image encryption. Chaotic signals are produced employing chaotic maps such as Logistic Map\cite{ausloos2006logistic},Lorenz Map\cite{sparrow1982lorenz},Henon Map\cite{benedicks1991dynamics},Tent Map\cite{li2017image}. Chaotic signals are extremely unpredictable and have characteristics that make them ideal for encryption purposes. With the use of chaotic signals in the proposed approach, randomness and initial condition sensitivity are greatly improved, rendering the encryption more resistant to statistical and brute-force attacks\cite{raghavendra2015robust},\cite{kumari2017survey}.

Diffusion methods have an important place in image encryption through the inclusion of randomness in pixel intensities that minimizes correlation between nearby pixels. In this method, the use of more than one diffusion level provides high security like 4-bit Diffusion\cite{li2024svdqunat},8-bit Diffusion\cite{li2021image},Radial Diffusion\cite{zhang2022color},Adjacent Diffusion\cite{wu2023diffumask}.These diffusions guarantee that a small modification of the original input image brings high changes to the encrypted image and thus high security against chosen-plaintext\cite{peng2006known} and differential attacks\cite{wu2011npcr}.

CAPTCHA (Completely Automated Public Turing test to tell Computers and Humans Apart)\cite{datta2005imagination} is widely employed to distinguish between human and automated users. In this new approach, a captcha image of random 6-character alphanumeric code against a coloured background autonomously generated by one of the algorithms as an authentication image. The captcha image passes through the same chaotic signal generation, pixel permuting, and diffusion process as the medical image.

The superimposition of the captcha and the medical image encrypted together constitutes the end encrypted output that can only be decrypted by those who have both decryption keys. The two-encryption mechanism is more secure by involving human authentication, and thus even if the encrypted medical image is intercepted, decryption without the captcha key is almost impossible\cite{kheshaifaty2023engineering}.

The suggested encryption scheme finds widespread applications in protecting medical images, serving a key role in secure storage and protection of sensitive health information. Within telemedicine frameworks, the scheme secures medical image transfer on public networks\cite{al2015crypto}, barring any illegal use or information theft. In the same vein, in Electronic Health Record (EHR) frameworks, the scheme guards images from storage against cyber attacks, keeping patient information confidential and secure\cite{cao2003medical}. In addition, it is also priceless in medical image repositories, where it protects vast patient databases in healthcare organizations\cite{andrade2012importance}. Remote diagnosis is also enhanced with this method, as it protects images relayed among healthcare providers and diagnostic facilities, allowing secure and trustworthy communication\cite{nurse2011trustworthy}. Cloud-based medical platforms are also enhanced with increased data confidentiality and integrity, which helps guarantee that data stored in cloud facilities remains safe from unauthorized parties\cite{shini2012cloud}. Utilizing chaotic signals and multi-level diffusion methods along with CAPTCHA-based authentication, the suggested method successfully solves the key problems of medical image encryption and decryption, ensuring a reliable solution for safe healthcare data management\cite{cao2017medical}.

The main contribution of this paper includes:

\begin{itemize}
    \item Medical imaging is an essential part of contemporary healthcare, providing detailed visualizations of anatomical structures using advanced imaging modalities. Of these, DICOM (Digital Imaging and Communications in Medicine) is the most common format because it not only stores the image data but also complete metadata with patient data, imaging parameters, and modality information. In contrast to traditional image formats such as PNG or JPEG, Our paper uses DICOM files which maintain the integrity and authenticity of medical information, and thus are essential in clinical diagnostics and analysis. This special characteristic of DICOM images introduces an extra layer of complexity to encryption since both the image data and related metadata must be protected securely.

\item Our work utilizes an hyperchaotic map, which greatly reinforces the security and strength of the encryption process. Hyperchaotic systems are characterized by unpredictable and complex behavior than conventional chaotic systems, making them more resilient to cryptographic attacks. The presence of multi-dimensional chaotic maps helps ensure that slight variations in input produce radically distinct encrypted outputs, hence reducing pattern recognition by adversaries to a large extent.

\item Along with chaos-based encryption, we use plethora of diffusion methods to effectively randomize the pixels of the medical images. This step greatly minimizes any remaining information that could give away the original content, so the encrypted result looks totally random and free from any recognizable patterns. The diffusion process is carefully crafted to optimize entropy and guarantee uniform pixel intensity distribution.

\item Our method is also concerned with the encryption of multimodality 3D volume rendering of anatomical models. In contrast to traditional 2D encryption methods, our technique is specifically designed to encrypt intricate 3D medical models without losing their volume integrity. This means that the encrypted 3D representations are secure for storage and transmission without sacrificing spatial and anatomical coherence required for correct medical analysis.

\item In addition, we present a novel method for encryption of certain regions of interest (ROI) of the 3D model, for eg; tumorous areas in brain images. The focused encryption facilitates safe processing and transmission of essential medical information while keeping the patient's identity and non-relevant anatomical areas confidential. Isolating sensitive areas and encrypting them individually allows us to reduce unwanted exposure of data and enhance privacy.

\item To add more protection to the data, we integrate image-based captcha generation as an additional layer of protection.The captcha generated consist of alphanumeric characters with special symbols and colored background along with some dots and lines to provide more complexity. The captcha is an additional factor of authentication that is a barrier to illegal entry or modification of the encrypted medical data. This combination of encryption and authentication enhances the security system as a whole and provides assurance that only authorized individuals are able to access medical data.
\end{itemize}
Our proposed method addresses the issues of encrypting complex medical information by combining advanced chaotic mapping, efficient pixel diffusion, multimodal 3D encryption, and secure authentication procedures. All these developments provide a highly secure and stable encryption system, which is critical for securing confidential medical information and patient identities in healthcare systems.

\section{Related Work}

Medical image encryption has become an essential aspect of secure healthcare data management, particularly in telemedicine and cloud-based medical systems. Various encryption methods and techniques have been proposed to ensure the confidentiality and integrity of medical data.

\subsection{Chaos-Based Medical Image Encryption}
Chaos-based encryption methods have gained significant attention due to their sensitivity to initial conditions and randomness, making them highly suitable for medical image protection. A study of chaos-based encryption system for DICOM medical images is mentioned in \cite{al2024chaos}, which utilizes chaos theory to enhance data security and prevent unauthorized access. The proposed system demonstrated robust performance against brute-force and statistical attacks. 

\subsection{Diffusion Techniques for Image Encryption}
Diffusion techniques are critical for altering pixel values to achieve high confusion and diffusion properties. A recent study on hybrid diffusion-based visual image encryption for secure cloud storage\cite{gao2024secure} introduced a multi-level diffusion strategy that spreads pixel values over a wider range. This method has shown to be effective against cryptographic attacks, especially in scenarios requiring long-term secure storage of medical data.

\subsection{Integration of Captcha for Authentication}
Captcha-based authentication mechanisms are widely adopted to ensure data integrity and restrict unauthorized access. A novel approach, QMedShield, integrates Captcha with quantum chaos-based encryption to secure medical images\cite{rajan2024qmedshield}. The scheme demonstrates high efficiency in authenticating users while maintaining robust image protection, making it suitable for secure telemedicine applications.

\subsection{Medical Image Security in IoT and Cloud Environments}
With the increasing adoption of IoT-enabled healthcare systems, protecting medical data during transmission and storage has become more challenging. A recent study introduced a quantum image encryption protocol designed for secure communication in IoT and cloud environments\cite{prajapat2025quantum}. This protocol ensures the confidentiality and integrity of medical data while minimizing computational overhead, making it feasible for real-time applications.

In summary, existing literature demonstrates significant progress in chaotic encryption, diffusion techniques, and Captcha integration for medical image security. Nevertheless, achieving a balance between security, computational efficiency, and usability continues to be an area requiring further investigation.

\section{Generation of extreme hyperchaotic signals}

Generation of extreme hyperchaotic signals is a multifaceted task where the task at hand is taking advantage of superior mathematical models so that one ends up generating incredibly dynamic and capricious behavior. Perhaps one of the brightest ways in which this is attained is via use of the 3D quadratic hyperchaotic maps, since these contain intensive dynamical behavior and generation of chaotic signals.

The 3D quadratic hyperchaotic map is formulated through the following set of equations:

\begin{equation}
\begin{aligned}
x_{n+1} &= a_1 x_n + a_2 y_n + a_3 y_n^2, \\
y_{n+1} &= b_1 - b_2 z_n, \\
z_{n+1} &= c x_n,
\end{aligned}
\label{eq:one}
\end{equation}

here, x, y, and z are the three state variables that change over discrete time steps n, while $a_1$, $a_2$, $a_3$, $b_1$, $b_2$, and $c$ stand in for the system parameters. The presence of all three positive Lyapunov exponents indicates that the 3D hyperchaotic map under study exhibits hyperchaotic behavior throughout a broad range of parameter space.

\begin{figure}[h!]
    \centering
    \includegraphics[width=1\textwidth]{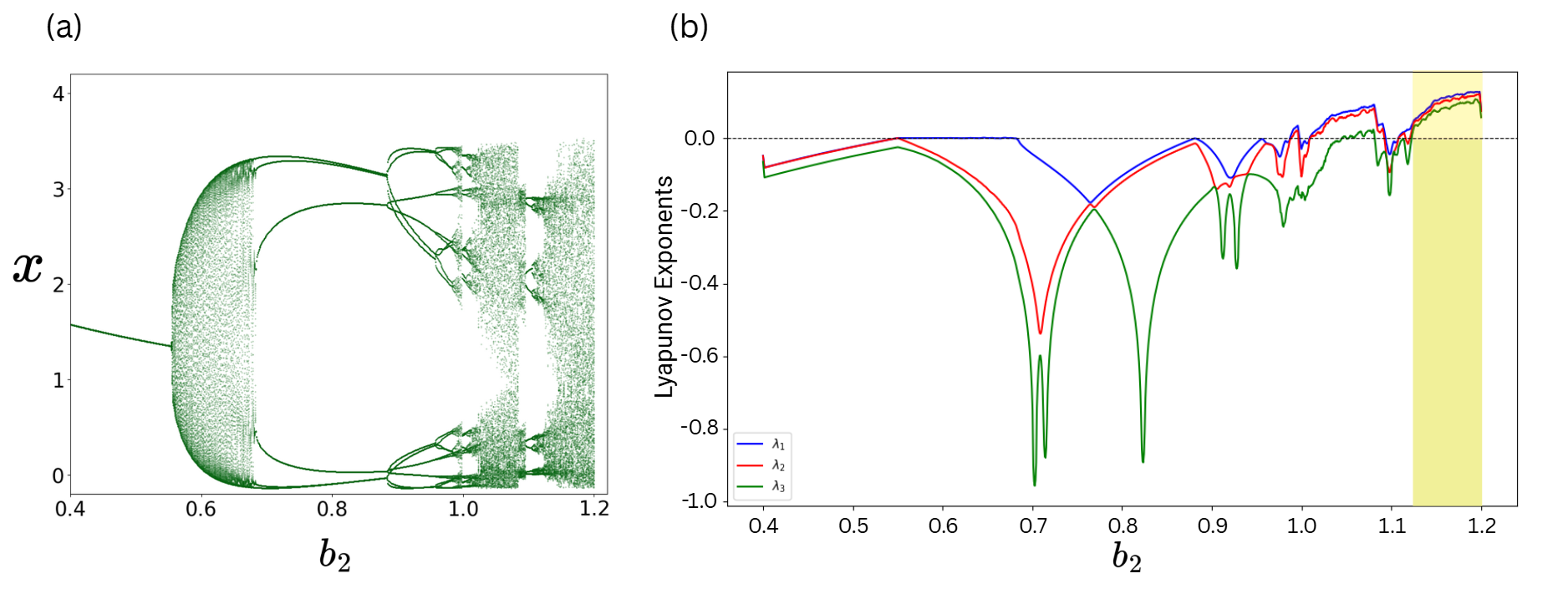} 
    \caption{Lyapunov exponent spectrum and one-parameter bifurcation diagram of 3D hyperchaotic map. The one-parameter Bifurcation diagram of a 3D hyperchaotic map is illustrated in Figure (a). The structural diagram of the above-mentioned 3D hyperchaotic map is provided in Figure (b), with the Lyapunov
exponent.}
    \label{fig:lyap} 
\end{figure}

In \cite{muni2024ergodic}, the extreme 3D hyperchaotic map was thoroughly analyzed theoretically and numerically, investigating the rich routes to hyperchaos. They revealed both ergodic and resonant torus doubling bifurcations, which were drastic changes in the dynamics of the system. Additional studies in\cite{muni2024pathways} identified numerous routes leading to hyperchaotic behavior, and parameter variation played a critical role.

\begin{figure}[h!]
    \centering
    \includegraphics[width=1\textwidth]{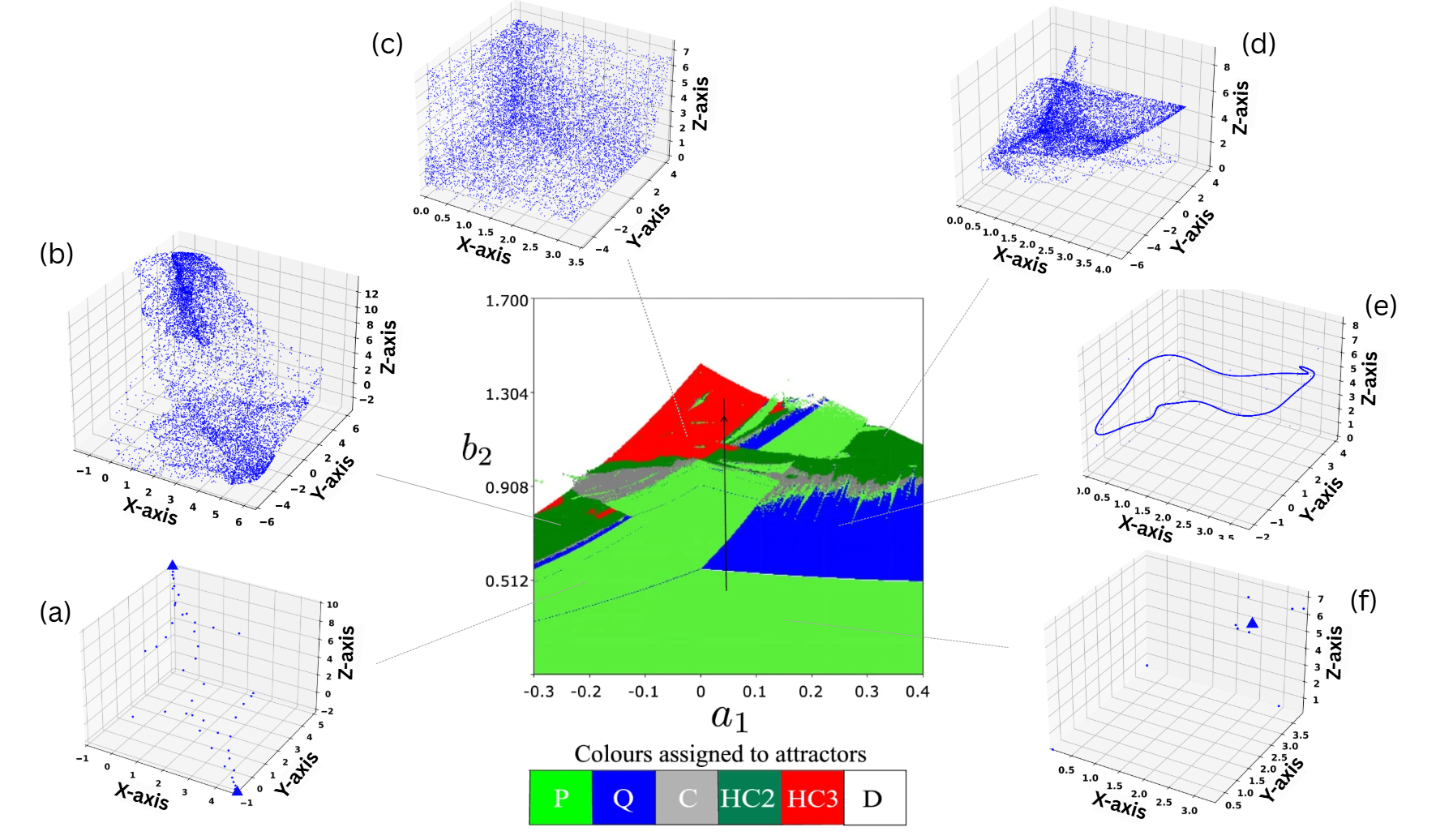} 
    \caption {A two parameter $a_1$-$b_2$ Lyapunov chart colored based on the legend for various attractors. The parameters are fixed as $a_2$=0.25,$a_3$=0.12,$b_1$=4,$c$=2.15}
    \label{fig:phase} 
\end{figure}

A one-parameter bifurcation diagram of $x$ vs $b_2$ is shown in Figure[\ref{fig:lyap}](a) and its corresponding three lyapunov exponents are shown in Figure[\ref{fig:lyap}](b) in red, blue and green. Observe that for lower values of b2,the 3D map equation \ref{eq:one} shows a steady state behavior. With increase in parameter $b_2$, the fixed point undergoes a Neimark Sacker bifurcation leading to the formation of a closed invariant curve with further increase in $b_2$, a Saddle node bifurcation on quasiperiodic closed invariant curve taken place leading to the formation of a stable period-six orbit. With subsequent increase in parameter $b_2$,it leads to period-doubling route to chaos and subsequently to the state of hyperchaos with the presence of all three lyapunov exponents with the presence of all three lyapunov exponents positive, see the yellow band region in Figure[\ref{fig:lyap}]

To gain a deeper insight into the rich dynamic of the 3D quadratic hyperchaotic map, the parameters $a_1$ and $b_2$ were simultaneously changed. The resultant Lyapunov exponent plot (refer [Fig\ref{fig:lyap}]) offers a three-parameter classification of the dynamic states, with regions representing:
\begin{itemize}

\item Periodic Orbits (P) - Represented in green, representing cycles that repeat.
\item Quasiperiodic Orbits (Q) - Represented in blue, depicting non-repeating bounded paths.
\item Chaotic Attractors (C) - Represented in grey, as an example of sensitive dependence upon initial conditions.
\item Hyperchaotic Attractors with Two Positive Lyapunov Exponents (HC2) - Represented in dark green, as a representation of increased complexity and more unpredictability.
\item Hyperchaotic Attractors with Three Positive Lyapunov Exponents (HC3) - Represented in red, for the most extreme kind of chaos.
\item Divergence (D) - Represented in white, for system instability and unbounded growth.
\end{itemize}

\subsection{Phase Portrait Analysis}

A useful method for studying dynamical systems, facilitating the determination of areas connected with quasiperiodicity, chaos, and hyperchaos, is the Lyapunov chart method. A two-parameter Lyapunov chart on a plane is represented in central part of Fig\ref{fig:phase}. To calculate Lyapunov exponents, we choose an orbit in an attractor and determine the exponents for its iterations with a standard technique\cite{benettin1980lyapunov}. For every point in the plane of a grid, we iterate the map from a random initial condition and calculate the Lyapunov spectrum. From the values of the Lyapunov exponents as indicated in Table\ref{tab:lyapunovt}, we color each point using the color palette at the bottom. The two-parameter Lyapunov chart plays a crucial role in determining the path to hyperchaos, particularly the regime with all three positive Lyapunov exponent

Some of the phase portraits are shown in Fig\ref{fig:phase}, highlighting the variety of behaviors obtained using different parameter settings:
\begin{itemize}
\item[(a)] At $a_1 = 0.15$, $b_2 = 0.12$, a \textit{period-2 bifurcation} is shown, which indicates a doubling of period as the system approaches chaos.

\item[(b)] At $a_1 = 0.25$, $b_2 = 0.75$, a \textit{hyperchaotic attractor} with two positive Lyapunov exponents is formed, showing greater complexity.

\item[(c)] At $a_1 = 0.35$, $b_2 = 1.15$, a \textit{hyperchaotic attractor} with all three positive Lyapunov exponents is seen, showing the maximum amount of hyperchaos with maximal unpredictability.

\item[(d)] At $a_1 = 0.05$, $b_2 = 1.2$, a distinct topology of a \textit{hyperchaotic attractor} with two positive Lyapunov exponents is reported, which is vastly different from that in (b).

\item[(e)] At $a_1 = -0.25$, $b_2 = 0.75$, a \textit{quasiperiodic closed invariant curve} is found, corresponding to stable but aperiodic oscillations.

\item[(f)] At $a_1 = -0.2$, $b_2 = 0.512$, the system stabilizes to a \textit{fixed point}. 

\end{itemize}

\begin{table}[h!]
\centering
\begin{tabular}{|p{4cm}|p{2.5cm}|p{4cm}|}
\hline
\textbf{Attractors} & \textbf{Abbreviations} & \textbf{Lyapunov exponents} \\
\hline
Periodic point attractor & P & $\lambda_i < 0, \quad i=1,2,3$ \\
\hline
Closed invariant curve & Q & $\lambda_1 = 0, \quad \lambda_i < 0, \quad i=2,3$ \\
\hline
Chaotic attractor & C & $\lambda_1 > 0, \quad \lambda_i < 0, \quad i=2,3$ \\
\hline
Hyperchaotic attractor with two positive Lyapunov Exponents & HC2 & $\lambda_3 < 0, \quad \lambda_i > 0, \quad i=1,2$ \\
\hline
Hyperchaotic attractor with three positive Lyapunov Exponents & HC3 & $\lambda_i > 0, \quad i=1,2,3$ \\
\hline
Divergent Trajectories & D & - \\
\hline
\end{tabular}
\caption{Attractors and their corresponding Lyapunov exponents.}
\label{tab:lyapunovt}
\end{table}

\section{Proposed Approach- Encryption}

\begin{figure}[htbp]
    \centering
    \hspace{-0.9cm} 
    \includegraphics[width=1\textwidth]{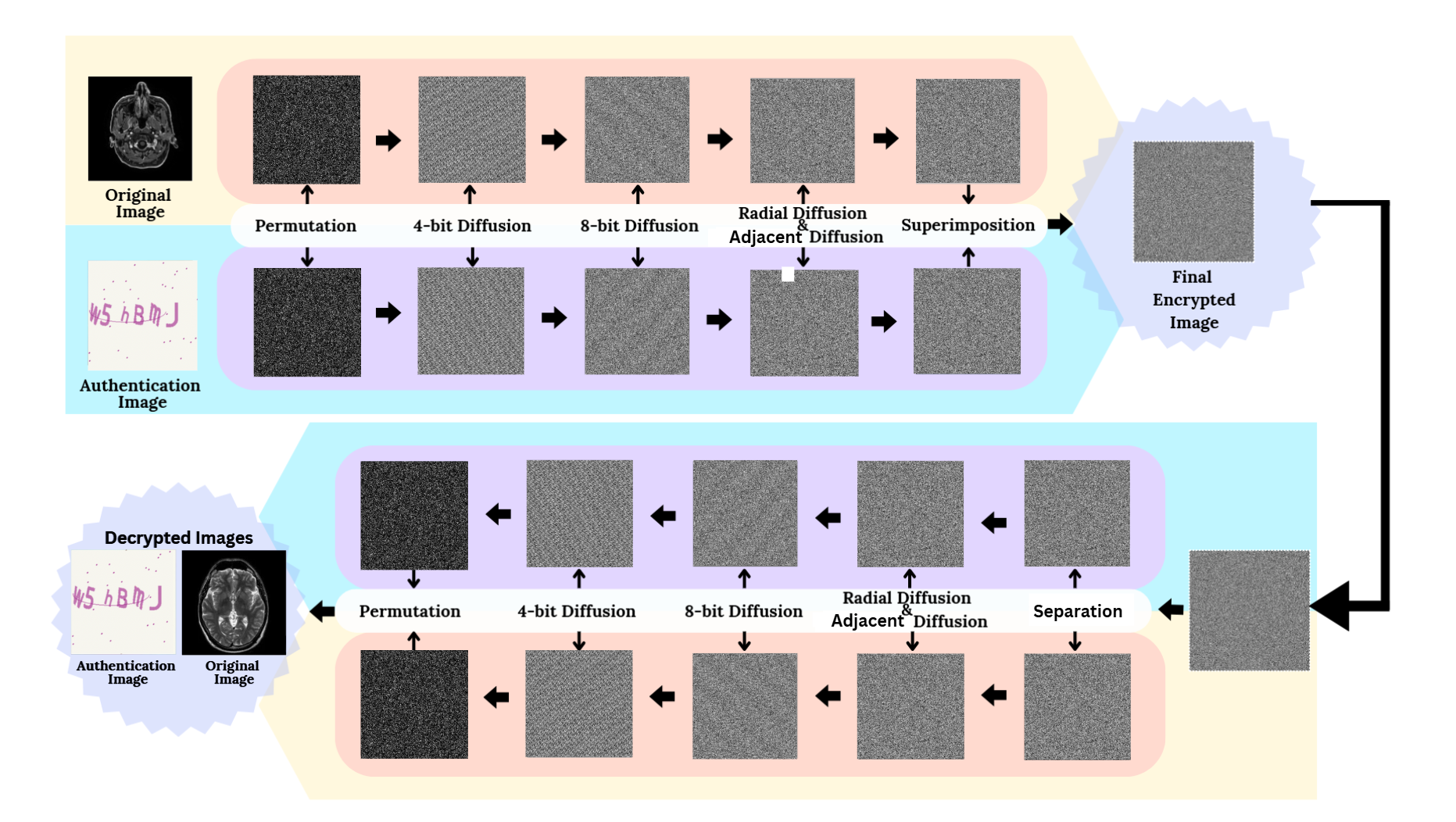}
    \caption{Flowchart showing every stage of the image encryption process, from obtaining input to producing a safe encrypted output. This guarantees that data protection procedures are visualized clearly.}
    \label{flowchart}
\end{figure}

The encryption of medical images starts with gathering high-quality images and preprocessing them via operations such as resizing and padding, which ensure standardization of their size for easy encryption. To enhance security, hyperchaotic signals introduces unpredictability, and pixel permutation randomizes the image organization. Sophisticated diffusion techniques such as 4-bit and 8-bit enhance complexity, rendering it difficult to track the original image. This layered approach secures medical information during storage and sharing as safe, private, and secure. The whole process is depicted in Figure \ref{flowchart}.

\subsection{Image Acquisition and Preprocessing}

Image acquisition and preprocessing are basic steps in medical image encryption to assure the quality and uniformity of input images prior to encryption. The main aim of this process is to receive high-quality images from different medical imaging modalities and carry out required preprocessing for improving the encryption process. Medical images can be obtained through a variety of imaging methods, such as Magnetic Resonance Imaging (MRI), which offers high-resolution images of soft tissues and internal organs; Computed Tomography (CT), which generates cross-sectional images with rich structural information; Ultrasound Imaging, which uses high-frequency sound waves to produce real-time images of internal body structures; X-ray Imaging, often employed for the detection of bone fractures and dental problems; Positron Emission Tomography (PET), which images metabolic processes in the body; and Digital Pathology Imaging, which takes microscopic images of tissue samples. These images are saved in DICOM (Digital Imaging and Communications in Medicine) format.

DICOM is the standard that is employed for medical image storage and transmission along with their related information. It helps to maintain compatibility and uniformity among various medical imaging equipment and healthcare systems and facilitates data integration and exchange smoothly. A DICOM image comprises two main components: the metadata and the image data. The image data holds the actual medical image obtained from modalities. The metadata, however, contains vital contextual information such as patient details (name, ID, age, sex, and medical history), image acquisition parameters (type of modality, scan date, resolution, and imaging parameters), image attributes (pixel spacing, orientation, grayscale values, and number of frames), device details (manufacturer, model, and calibration), and study and series details (study ID, series number, and image sequence). The importance of DICOM metadata is that it can guarantee precise patient identification, ensure consistency between scans, aid in image management and retrieval, enable automated analysis, and enhance data integrity and protection. Maintaining the accuracy of metadata during encryption of medical images is vital, as any loss or alteration can cause grave errors in patient treatment and data analysis.

\begin{figure}[h!]
    \centering
    \fontsize{10}{12}\selectfont   
    \normalsize                    
    \includegraphics[width=1\linewidth]{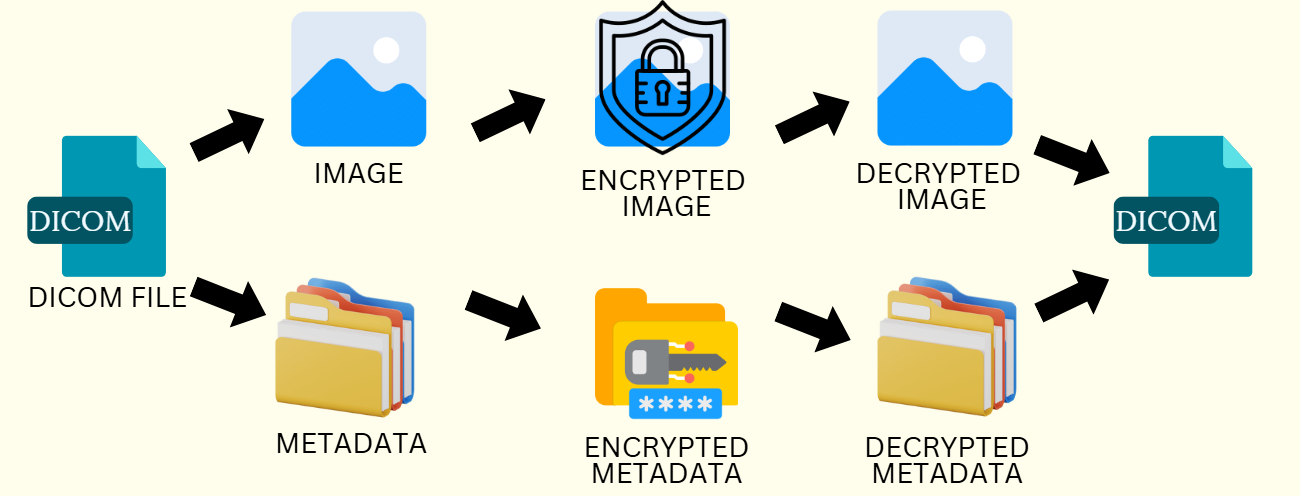}
    \caption{ Flowchart showing how the dicom file is being handled throughout the encryption and decryption processes.
}
    \label{figure:dico}
\end{figure}

Keeping the metadata accurate throughout encryption of medical images is crucial, as any loss or modification can lead to serious errors in patient treatment and data analysis. To maintain both security and integrity, the DICOM file is divided into two parts, the image data and the metadata before being encrypted. The image is then processed with sophisticated encryption via different diffusion methods, in addition to hyperchaotic map signals that unpredictably reassign pixel locations to maximize unpredictability and security. At the same time, the metadata is encrypted safely to maintain its sensitive data. At decryption, this operation is reversed, both the image and metadata are decrypted correctly, and the original DICOM file is rebuilt without loss. This method ensures that both the medical image and its related metadata remain confidential, tamper-proof, and recoverable, thus enabling safe and secure medical diagnostics and data exchange,see figure \ref{figure:dico}.

Preprocessing plays a vital role in the process of preparing raw medical images for enhancing encryption efficiency and accuracy. The most frequently employed preprocessing operation is image resizing and cropping, which is done by modifying the image size to conform to the input size specification of the encryption algorithm. Not only does this operation normalize the image size but also minimizes computational complexity in the encryption process.

\subsection{Padding and Hyperchaotic Signal Generation}
Padding is a critical preprocessing operation in encrypting medical images, with the main purpose being to achieve image size uniformity before the encryption operation. Medical images obtained from various imaging devices can be quite dissimilar in their size, which creates difficulties in applying consistent encryption and processing. Padding achieves standardization of the image size and brings it into alignment with the encryption algorithm and minimizes boundary effects during the chaotic mappings.

\begin{figure}[h!]
    \centering
    \fontsize{10}{12}\selectfont   
    \normalsize                    
    \includegraphics[width=1\linewidth]{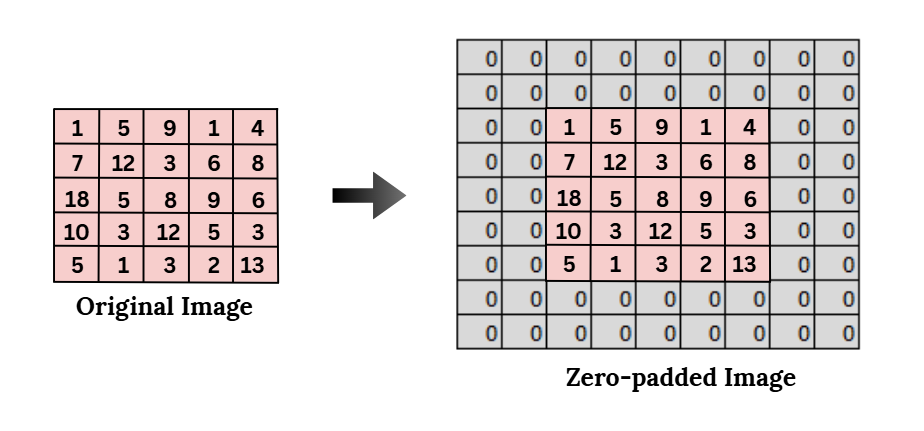}
    \caption{Zero-padding is the used to add additional rows and columns with zeros around an image's boundaries. By doing this, the picture size is maintained during processes like convolution.
}
    \label{figure:zero}
\end{figure}

In this encryption scheme, padding is done to increase the image size to the nearest power of two, which maximizes the efficiency of chaotic signal generation and diffusion methods. Zero padding and mirror padding are popular methods. Zero padding adds zeros to the borders of the image, whereas mirror padding copies the edge pixels to ensure a natural boundary. Here we are using Zero padding as shown in figure \ref{figure:zero}. The padded image is turned into a square matrix to enable uniform processing and application of chaotic maps. This ensures that the encryption scheme maintains its efficiency and robustness irrespective of the original image size.

Hyperchaotic signal generation is a core aspect of the designed encryption algorithm, which plays an important role in making the system complex and secure. Hyperchaotic systems demonstrate high-dimensional chaotic behavior, having more than one positive Lyapunov exponents. Hyperchaotic systems offer better unpredictability and robustness against usual cryptographic attacks as compared to usual chaotic maps.

\begin{algorithm}
\caption{Chaotic Pixel Permutation}
\textbf{Input:} Medical Image \( I \) (DICOM) \\
\textbf{Output:} Permuted Image \( I_{perm} \)

\begin{algorithmic}[1]
    \State Extract DICOM metadata and pixel data.
    \State Zero-pad \( I \) to square matrix \( I_{padded} \).
    \State Generate chaotic sequences \( S_x, S_y \).
    \State Permute rows using \( S_x \), columns using \( S_y \).
    \State Output permuted image \( I_{perm} \).
\end{algorithmic}
\end{algorithm}

\subsection{Pixel Random Permutation}
Pixel random permutation breaks the spatial order of the image matrix, providing an added layer of protection. This operation employs the previously generated chaotic sequence to establish a new pixel ordering. By permuting the pixel locations, the permutation masks the original structure of the image, and it is not easy to infer any useful information.

The permutation algorithm works by relocating every pixel to a new location using the chaotic sequence. A pixel, for instance, initially found at position (x, y) can be relocated to position (x', y') within the permuted matrix. This permutation is deterministic but is perceived as random to anybody without the access to the chaotic key with which the sequence has been produced.

This operation makes it so that even if an attacker achieves a partial access of the encrypted image, due to the absence of spatial coherence, it is extremely difficult for him to rebuild the original image. The permutation also supports the chaotic embedding operation since the two operations combined produce a highly disordered and safe image matrix.

\begin{figure}[h!]
    \centering
    \fontsize{10}{12}\selectfont   
    \normalsize                    
    \includegraphics[width=1\linewidth]{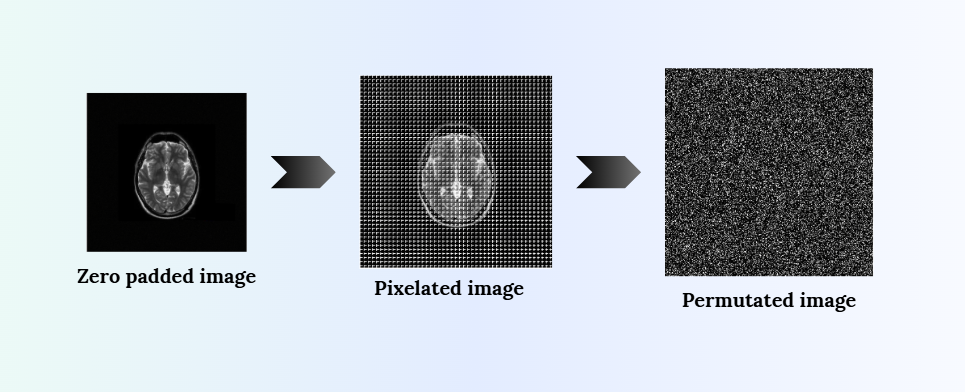}
    \caption{Conversion of Zero-Padded Image to a Pixel Random Permutated Image for Improved Security and Data Augmentation.
}
    \label{figure:prpi}
\end{figure}

Pixel reordering is very effective when it is used with subsequent diffusion phases. Shattering the spatial connection between pixels ensures that the effects of diffusion are spread uniformly across the image. This use of techniques increases the complexity and immunity of the encryption to attacks.

The result of this phase is a mixed matrix whose pixel structure has nothing to do with the original picture as shown in figure \ref{figure:prpi}. This jumbled matrix is used as an input for the diffusion process in order to provide the encryption system with a maximum degree of security.

\subsection{2-bit Diffusion}
2-bit diffusion adds an intermediate level of complexity to the encryption process. Every pixel is split into several 2-bit segments, and the respective segments of the chaotic key are utilized for segment-wise XOR operations. This approach adds more confusion and helps in the diffusion process, without incurring a very high computational overhead.

For example, a pixel value of 10101100 can be divided into 10, 10, 11, 00 and XORed with corresponding 2-bit parts of the key. This guarantees that even minor variations in the pixel or key result in perceptible variations in the ciphertext, strengthening the security via the avalanche effect(refer figure \ref{figure:bitencryp}a).

\subsection{4-bit Diffusion}

4-bit diffusion increases the process complexity further by segmenting each pixel into 4-bit segments. It raises the number of
possible combinations and improves the effect of diffusion which makes it more difficult for attackers to understand
the relationship between the plaintext and the ciphertext.
Every pixel is segregated into 4-bit segments here, and so is the key. Segment-wise XOR operations are performed,
such that each 4-bit unit of the pixel interacts with the corresponding key unit.

\begin{figure}[h!]
    \centering
    \fontsize{10}{12}\selectfont   
    \normalsize                    
    \includegraphics[width=1\linewidth]{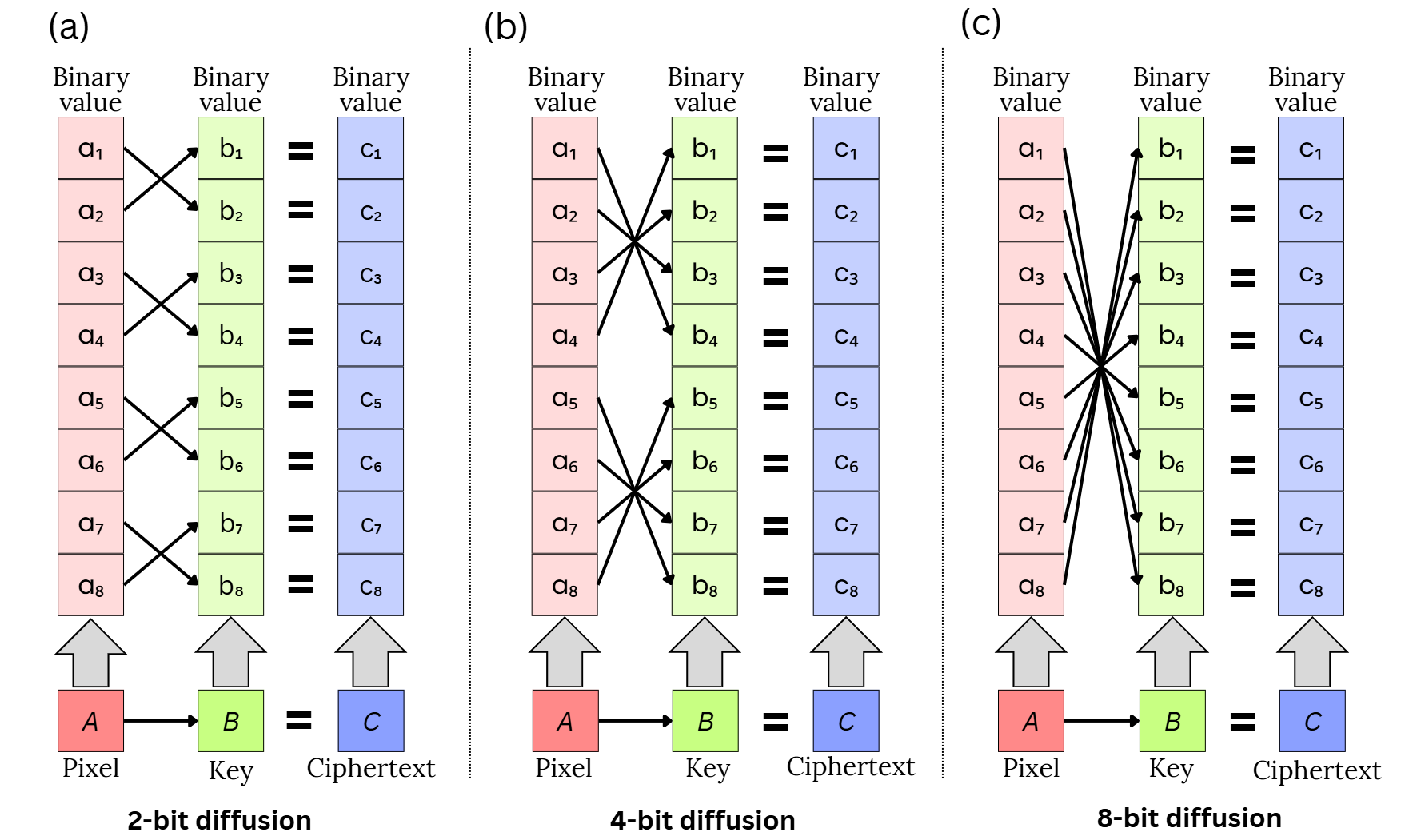}
    \caption{ 2-bit, 4-bit, 8-bit diffusion of pixels 
}
    \label{figure:bitencryp}
\end{figure}

The ciphertext is produced as:
The extra bits per segment provide enhanced diffusion as each segment has more information. For example, a pixel value
of "11001010" and a key of "10100101" will produce a ciphertext with higher entropy. The application of 4-bit diffusion
provides interdependence among bits within a segment as shown in figure \ref{figure:bitencryp} (b).

This interdependence amplifies the avalanche effect – a single-bit change in the input can lead to significant alterations in the output, enhancing encryption strength.
Moreover, the 4-bit configuration is computationally efficient, striking a balance between security and processing time. It is particularly suitable for lightweight encryption applications where resource constraints are a concern.

\begin{algorithm}
\caption{Hyperchaotic Bitwise Diffusion}
\textbf{Input:} Permuted Image \( I_{perm} \)\\
\textbf{Output:} Diffused Image \( I_{diffused} \)

\begin{algorithmic}[1]
    \State Generate chaotic sequence \( S_z \).
    \State For each pixel, segment into 2-bit / 4-bit / 8-bit units.
    \State Perform segment-wise XOR with corresponding key segments from \( S_z \).
    \State Apply adjacent row and column XOR diffusion.
    \State Apply radial XOR diffusion using \( S_z \).
    \State Output the diffused image \( I_{diffused} \).
\end{algorithmic}
\end{algorithm}

\subsection{8-bit Diffusion}

8-bit diffusion is the strongest implementation of the three. A single 8-bit unit is a pixel, and the whole
byte is intermingled with the respective key byte. Thus, there is full diffusion across all bits and maximum security from
statistical and brute-force attacks.
Pixel and key are both considered to be 8-bit units. XOR operations are directly applied over these units in order to yield the ciphertext.

This approach ensures that every bit of the pixel contributes to the output, and any change in the key or pixel propagates across
the entire byte. For example, a pixel value of “11110000” and a key of “00001111” would result in a ciphertext that is highly
unpredictable, refer figure \ref{figure:bitencryp}(c). 8-bit diffusion is the maximum avalanche effect that makes it very hard to discover correlations between output and input by attackers. The wider bit also adds higher entropy, which drops the chances of effective attacks.

Its computational overhead is still superior than that of two-bit or 4-bit settings. This notwithstanding, its strong
security features render it the preferred choice for packages with high levels of encryption needs, including stable communications
and virtual watermarking

\subsection{Adjacent Diffusion}

The subsequent diffusion process is the final step of encryption in the suggested system. It serves to provide thorough clarity of the input data with row and column propagation operations. This process involves a bitwise XOR operation to create the final code where the value Each pixel is in adjacent rows and columns. which is impacted Including these interactions Adjacent propagation
steps add further complexity to the encoding. This renders data extremely resistant to rebuilding or unauthorized access.
The process starts with row-wise neighboring diffusion. In this operation, every row in the matrix is refreshed by doing a bitwise
XOR between its pixel values and the nearby pixel values in the following row. This produces a cascading
effect whereby the encryption of a row relies on data from the row below, disseminating the encryption effect vertically throughout the
matrix. For the bottom row in the matrix, the values are not changed since there are no rows beneath it.

The output of this process is an
intermediate matrix which has a vertical diffusion, strengthening the inter-row dependability of the pixel values.
After row-wise operation, the matrix goes through column-wise adjacent diffusion. Just like in row-wise diffusion, this
step processes column by column, with each column's pixel values being updated through a bitwise XOR operation with
the pixel values of the column to its right. This step spreads the encryption horizontally so that the neighboring columns also
control the final values. The last column is not altered since there are no columns to its right. The combination of vertical and
horizontal diffusion will make the final matrix completely encrypted, with every pixel's value relying on its neighboring neighbors in
both dimensions.
The product of the adjacent diffusion process is the resulting diffused matrix, which is heavily scrambled and obscured from
the input matrix.

Visualization of the process reveals the change at each step: the original matrix, the output after row-wise
diffusion, and the resulting matrix after column-wise diffusion. Such visual presentations identify the strength of the adjacent
diffusion step in fully changing the appearance of the input data.
Through the integration of row-wise and column-wise operations, the adjacent diffusion step is the last layer of encryption in the proposed system. The step strongly secures data by making the encrypted matrix both computationally difficult to reverse and resistant to cryptanalysis. The adjacent diffusion process finalizes the encryption pipeline, generating a secure and strong representation of the original data.

\subsection{Radial Diffusion}

Radial diffusion augments neighboring diffusion by adding a circular propagation method that further makes decryption more challenging. Under this process, the diffusion occurs starting from the center and extending outward in the form of concentric circles. Pixels in each circle are diffused using bitwise XOR operations among neighboring pixels radially. This method guarantees that data from the center area affects the outer layers, thus an encryption pattern radiating outward is established.

Radial diffusion not only makes the encryption more complex but also adds security with multidirectional propagation. The interaction of circular and neighboring propagation complicates it significantly for attackers to decode the original information. Radial diffusion wraps up by merging the outer circle with neighboring diffused rows and columns to produce a multi-layered, completely diffused encryption pattern.

The combined use of adjacent and radial diffusion processes renders the encryption process extremely cryptanalysis-resistant. The cascading action of adjacent diffusion and radial expansion of pixel values creates an invulnerable and well-scrambled output matrix. Combinations of radial and adjacent diffusion constitute the final layer of security, which guards the data presentation against unauthorized manipulation.

\section{Captcha-Based Authentication Image Generation}

Captcha-based authentication is important in enforcing the security of medical image encryption to guarantee that only legitimate users can retrieve the encrypted information. In our proposed scheme, we employ Captcha as an authentication image to provide an extra security layer. Captcha (Completely Automated Public Turing test to tell Computers and Humans Apart) defends against automatic brute-force attacks by telling apart human users from automated bots. The Captcha images are generated dynamically with random alphanumeric characters, which are transformed using rotation, scaling, warping, and adding noise to make them immune to optical character recognition (OCR) by robots. The Captcha images are encrypted with the same hyperchaotic signal and diffusion methods applied to the medical images, and the resultant encrypted image is obtained by overlaying the encrypted Captcha over the encrypted medical image. At decryption, accurate Captcha entry is required to authenticate users, so that the system is secure even if the encrypted image is intercepted. This integration greatly strengthens data security, especially in telemedicine and remote diagnostics, by introducing a strong human-verification layer to the encryption process.

\subsection{CAPTCHA Encryption Workflow}

Padding is a required process in normalizing the size of medical images prior to encryption, to provide consistency regardless of differing dimensions among imaging modalities. By padding the original image to the next power of two without changing the aspect ratio, it introduces minimal redundancy while normalizing dimensions for more effective noisy signal generation and pixel reordering. Standardization enhances encryption accuracy and reduces distortion during decryption. Central to encryption is the process of generating hyperchaotic signals, bringing with it high unpredictability and sensitivity. The application of three-dimensional hyperchaotic sequences in cubic parameter-based and small random numbers-seeded sine-based chaotic functions guarantees the establishment of unpredictable and complex patterns for signals that amplify encryption security. These sequences facilitate pixel random permutation, distrupting neighbor relationships between pixels using a permutation matrix to cut statistics correlation as well as to conceal visual perception. Post-permutation, 4-bit and 8-bit diffusion processes expand randomness by incorporating XOR and byte rotation at the byte and bit level, so that even minor alterations in the original image yield highly dissimilar encrypted output. To remove the last of pixel correlation, neighboring and radial diffusion methods are employed. Adjacent diffusion introduces pixel-level randomness by affecting each pixel in relation to its neighbors, and radial diffusion breaks spatial coherence by diffusing pixels in circular orders from the image center, both driven by chaotic signals. Combining these steps provides a stable encryption system that withstands multiple types of cryptographic attacks such as known-plaintext and chosen-plaintext attacks.

\begin{algorithm}
\caption{CAPTCHA Superimposition and Final Encryption}
\textbf{Input:} Diffused Image \( I_{diffused} \), CAPTCHA Image \( C \) \\
\textbf{Output:} Final Encrypted Image \( I_{enc} \)

\begin{algorithmic}[1]
    \State Encrypt CAPTCHA \( C \) using the same permutation and diffusion steps.
    \State Perform XOR between \( I_{diffused} \) and \( C_{enc} \).
    \State Embed encrypted metadata into DICOM format.
    \State Output encrypted image \( I_{enc} \).
\end{algorithmic}
\end{algorithm}

\section{Superimposition of Encrypted Images}

\begin{figure}[h!]
    \centering
    \fontsize{10}{12}\selectfont   
    \normalsize                    
    \includegraphics[width=1\linewidth]{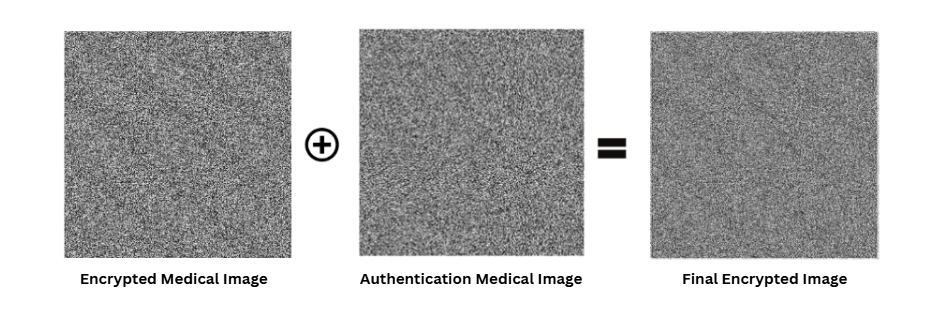}
    \caption{ Superimposition of encrypted medical image and the authentication image, ensuring secure verification and data integrity. A seamless fusion of encryption and authentication for robust medical data protection.
}
    \label{figure:superimposition}
\end{figure}

Superimposition of encrypted images is a sophisticated method that improves data security through the amalgamation of various encrypted images into one composite image, refer figure \ref{figure:superimposition}. This process is such that even if an encrypted image is tampered with, obtaining useful information becomes highly impossible. In the process, there is generation of several encrypted images from hyperchaotic signals and each signal representing a different medical image.

The superimposition is carried out pixel-wise over the encrypted images with bitwise XOR-based mathematical operations. Chaotic keys and permutation matrices employed at encryption are specific to every image, making the security even stronger. Not only does this increase the complexity of the ciphertext but also provides greater immunity to brute-force and differential attacks.

Superimposed encrypted images maintain the structural randomness of each individual encryptions while reducing data redundancy. It is especially beneficial in the case of the transmission of numerous medical images via insecure channels, where this technique reduces data to a single, secured file. In decryption, the same process is reversed by the separation of images using the corresponding chaotic keys and orders of permutations, returning the original images without any loss of data.

By utilizing superimposition, our scheme attains a multi-level security strategy that renders it highly immune to statistical and cryptanalytic attacks.

\section{Decryption}

Decryption is the inverse of encryption, whose purpose is to recover the original medical image from the encrypted one. Decryption uses the exact same sequence of actions as encryption but in reverse order using the same chaotic keys and permutation matrices as those used in encryption.

\begin{figure}[h!]
    \centering
    \fontsize{10}{12}\selectfont   
    \normalsize                    
    \includegraphics[width=1\linewidth]{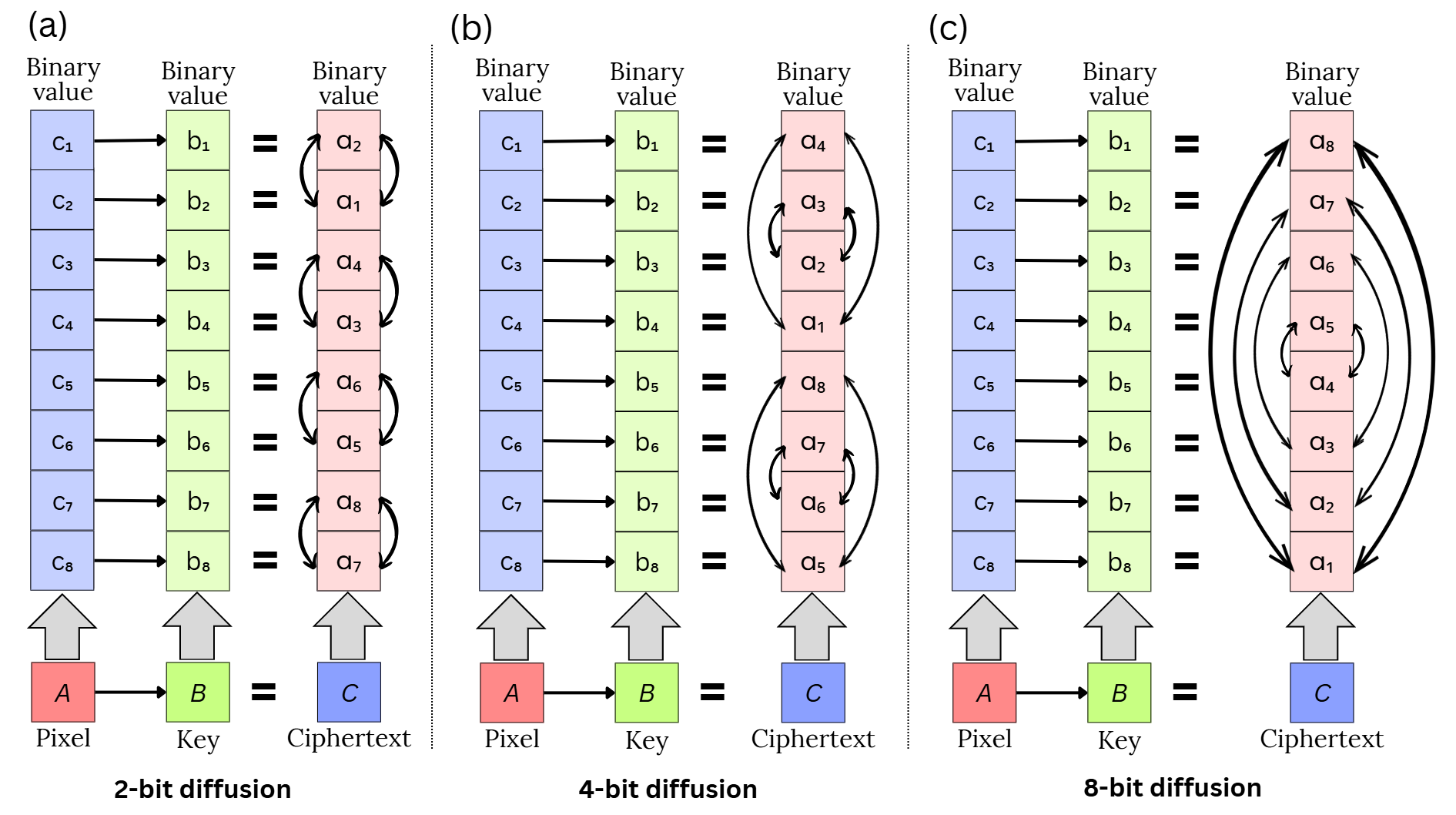}
    \caption{ Decryption process involving the reversing bitwise diffusion techniques
}
    \label{figure:bitdecr}
\end{figure}

The method starts with recovering the superimposed encrypted image and de-superimposing to get individual encrypted images. After separation, each encrypted image passes through inverse diffusion processes(refer figure \ref{figure:bitdecr}), both radial and adjacent diffusion reversal, to cancel the impact of chaotic mixing in the process of encryption.

Then, the random permutation of the pixels is reversed by using the inverse of the random permutation matrix, thereby recovering the pixel locations to their initial arrangement. The equations for generating the chaotic signals are followed again with the same initial parameters in order to reproduce the chaotic sequences employed in the encryption process exactly. These sequences are utilized to decrypt the pixel values by undoing the hyperchaotic transformations that were done in the encryption process.

\begin{figure}[h!]
    \centering
    \fontsize{10}{12}\selectfont   
    \normalsize                    
    \includegraphics[width=1\linewidth]{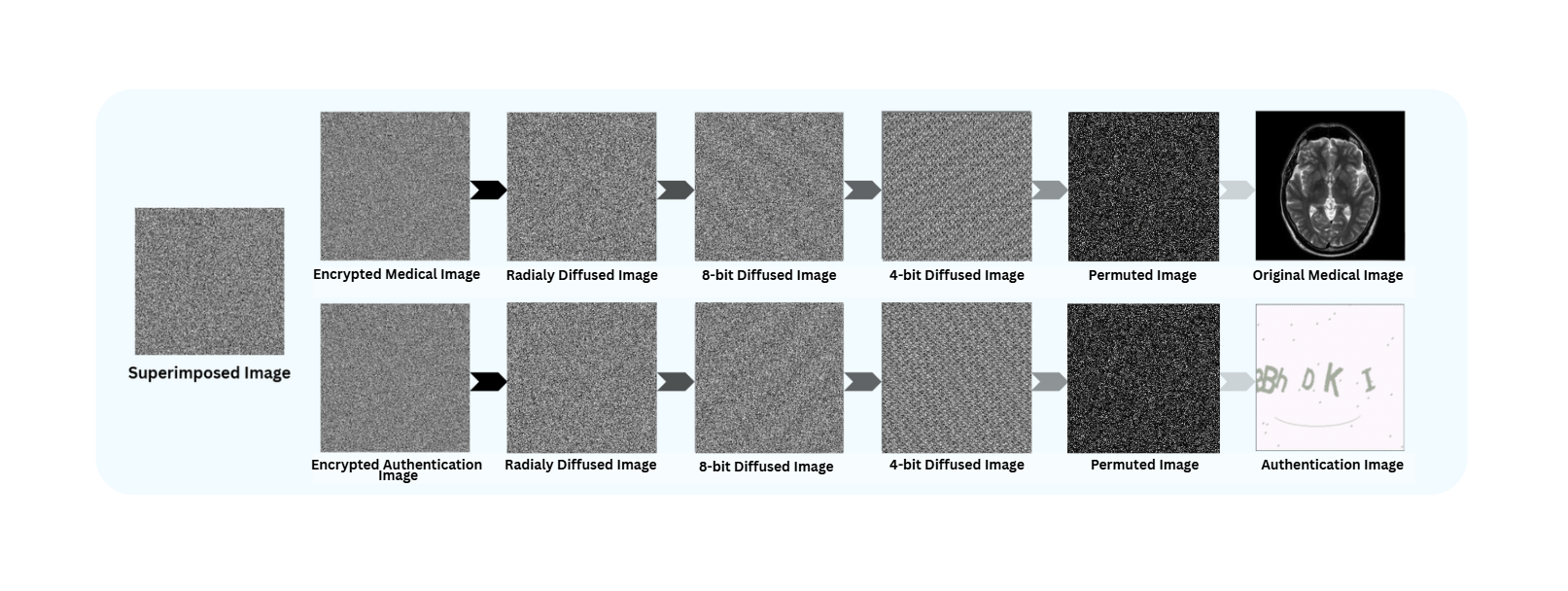}
    \caption{ Decryption process involving the reversing of all the steps done at the time of encryption
}
    \label{figure:decru}
\end{figure}

Finally, padding inserted at the encryption time is removed and the decrypted image is restored to the original input size as shown in figure \ref{figure:decru}. CAPTCHA-based authenticating image, is checked for integrity and authenticity of the decrypted result.

By exactly executing the decryption procedures, the original medical image can be restored in a non-distorted and non-lost form, retaining its diagnostic quality. The correctness and stability of the decryption procedure are essential for guaranteeing that medical experts can reliably read the decrypted images.

\begin{algorithm}
\caption{Decryption of Hyperchaotic Encrypted Medical Image}
\textbf{Input:} Encrypted Image \( I_{enc} \) \\
\textbf{Output:} Decrypted Image \( I_{orig} \)

\begin{algorithmic}[1]
    \State Extract encrypted pixel data and metadata.
    \State Regenerate chaotic sequences \( S_x, S_y, S_z \).

    \State Separate and decrypt CAPTCHA \( C_{enc} \), verify authenticity.

    \State Apply inverse radial diffusion using \( S_z \).
    \State Apply inverse adjacent diffusion (rows and columns).

    \State XOR with \( S_z \) for inverse 8-bit, 4-bit and 2-bit diffusion.

    \State Unshuffle rows and columns using inverse of \( S_x \) and \( S_y \).

    \State Remove padding and restore original image size.
    \State Return decrypted image \( I_{orig} \).
\end{algorithmic}
\end{algorithm}

\section{Encryption of Multimodel application}

This encryption algorithm is versatile and can be used in different medical imaging modalities. As a result of the strength in chaotic signal production and diffusion, the algorithm has the capability to secure images retrieved from different sources such as MRI, ultrasound, X-ray, and digital pathology images.

\begin{figure}[h!]
    \centering
    \fontsize{10}{12}\selectfont   
    \normalsize                    
    \includegraphics[width=1\linewidth]{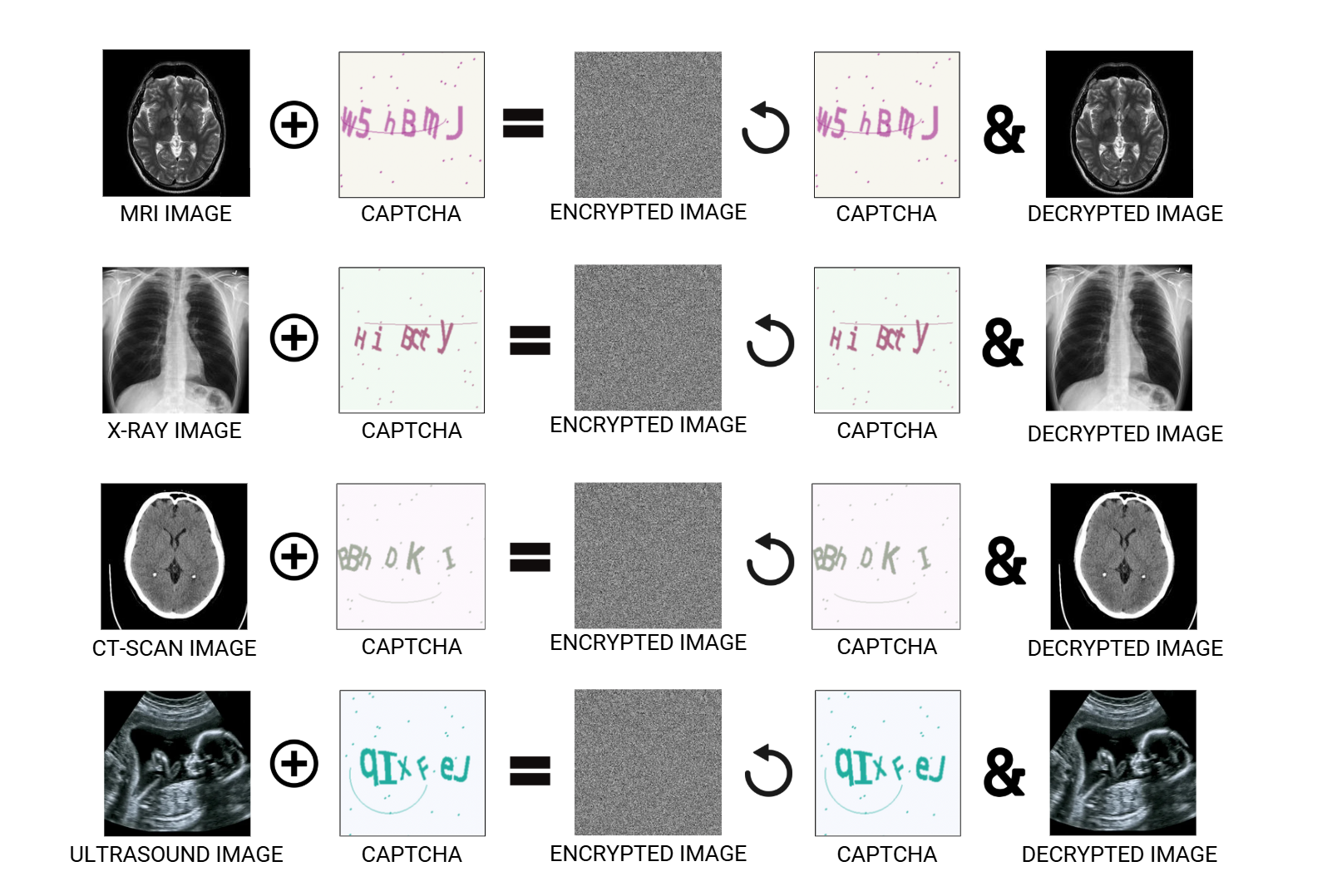}
    \caption{ The pathway illustrating the encryption and decryption of various medical images using a robust encryption algorithm. Demonstrates multimodality through secure and efficient data handling.
}
    \label{figure:multi}
\end{figure}

MRI generates detailed images of internal organs and soft tissues, and ultrasound imaging records real-time internal structure visualizations with high-frequency sound waves. X-ray imaging is used typically for the identification of bone fractures and dental conditions, and digital pathology imaging yields microscopic images of tissue samples. In spite of the inherent image differences in resolution, contrast, and texture, the encryption algorithm accommodates these differences with ease. In the figure\ref{figure:multi} we can see multiple images being encrypted and decrypted. 

The preprocessing phase of resizing, cropping, and denoising makes the system consistent irrespective of the input modality. Additionally, the dynamic characteristic of hyperchaotic signal creation and the versatility of diffusion methods allow for accepting various types of images without weakening security.

By enabling the encryption of various medical imaging modalities, the presented encryption algorithm proves feasible in numerous different healthcare applications. Such broad applicability is essential to provide effective end-to-end data security in state-of-the-art medical systems that support various image formats and types.

\section{Encryption of 3D Model}

\begin{figure}[h!]
    \centering
    \fontsize{5}{6}\selectfont   
    \normalsize                    
    \includegraphics[width=0.5\linewidth]{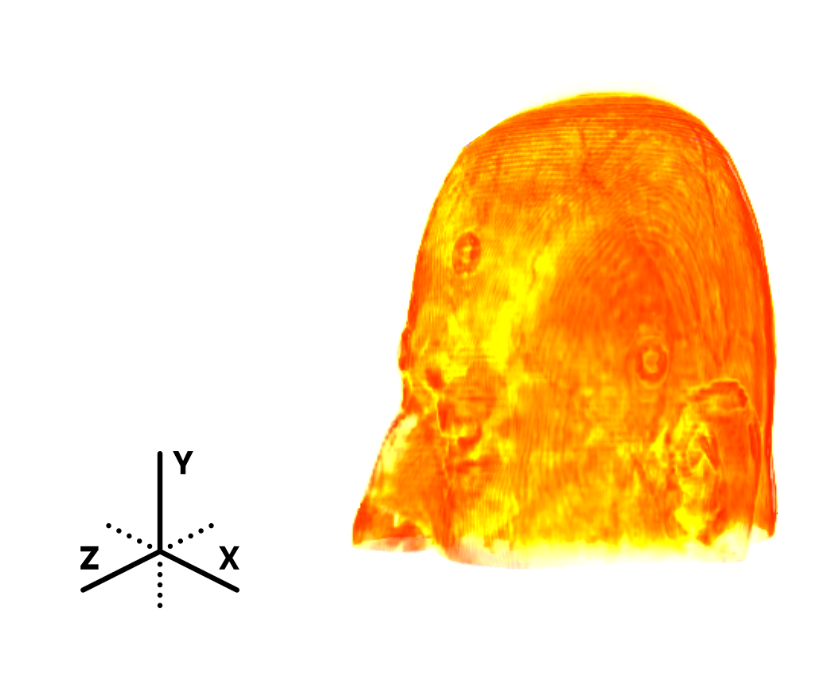}
    \caption{ Volume rendering obtained from the medical scan utilizing the DICOM tags. Clearly and precisely visualizing complex data.
}
    \label{figure:3dvol}
\end{figure}

Encrypting 3D medical models is challenging because they have a complicated structure and are high-dimensional in nature\ref{figure:3dvol}. To provide complete protection without losing flexibility, our new encryption algorithm has two different modes: Whole Model Encryption and Partial Model Encryption.

\subsection{Whole Model Encryption}
\begin{figure}[h!]
    \centering
    \fontsize{5}{6}\selectfont   
    \normalsize                    
    \includegraphics[width=0.5\linewidth]{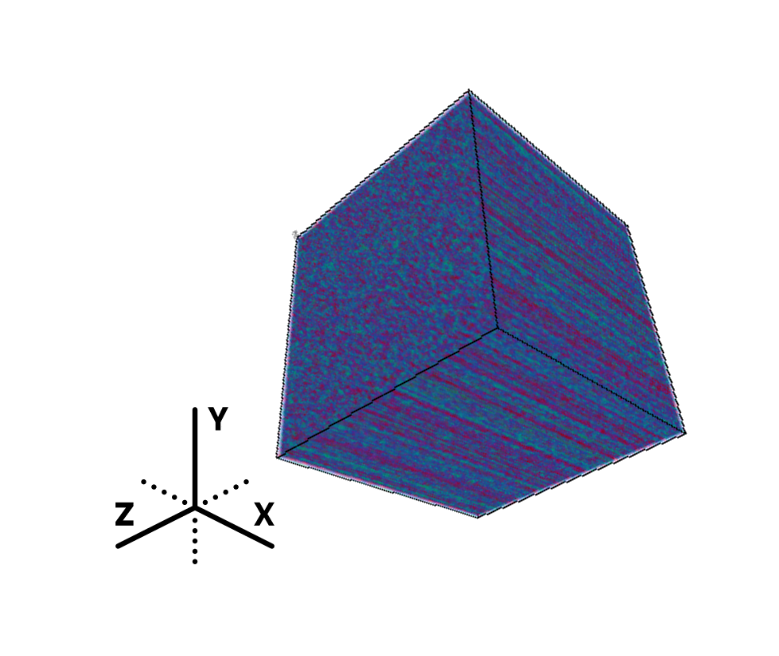}
    \caption{A 3D encrypted model that expertly blends every layer into a single, cohesive whole. ensuring complete data protection and strong security..
}
    \label{figure:3dwhole}
\end{figure}

The medical images obtained from MRI scan are stacked together to generate the 3D model of the patient head. Figure\ref{figure:3dvol} shows the 3D model generated using several DICOM tags like pixel spacing, slice thickness, image orientation etc. In whole model encryption, entire 3D model of the medical scan is encrypted. Our method is capable of encryption of each slice of any series like axis, coronal or seggital Finally, algorithm creates the 3D encrpted cube as shown in figure\ref{figure:3dwhole}. The cube can be decrypted to recover original 3D model. The method is specifically designed to prevent data leakage and identity theft of patient during the circulation of medical scans for research, education, and publishing. This method is especially effective for delicate medical procedures where the entire structure must be protected, like whole-body CT or MRI scanning.

\begin{figure}[h!]
    \centering
    \fontsize{5}{6}\selectfont   
    \normalsize                    
    \includegraphics[width=0.5\linewidth]{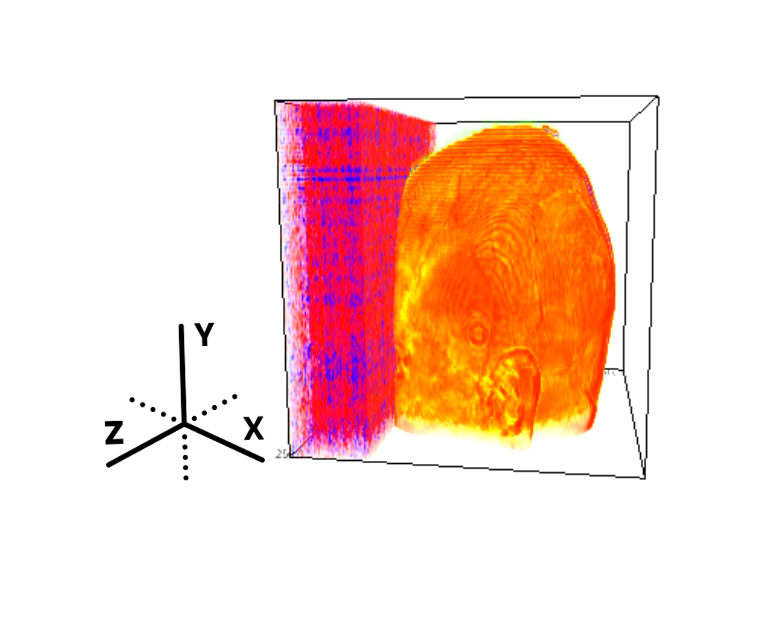}
    \caption{ A three-dimensional encrypted model that combines partial encryption with all levels for maximum security. effortlessly strikes a balance between performance and protection..
}
    \label{figure:3dpart}
\end{figure}

\subsection{Partial Model Encryption}
Partial model encryption, however, offers a more specific solution by enabling selective encryption of certain areas of interest in the 3D model. For example, if the face area of a head scan must be kept safe for privacy purposes, the algorithm can be made to encrypt just that area and leave the remaining part of the model intact as shown in Fig\ref{figure:3dpart}. In this way, there is considerable computational overhead reduction and secure sharing of non-sensitive parts while ensuring confidentiality for sensitive areas.

By providing the whole and partial model encryption models, our algorithm is flexible in the level of securing 3D medical data according to the needs. The dual treatment guarantees that health practitioners can select the most suitable encryption method without compromising on data integrity or security.

\section{Security Analysis}

One of the most critical methods to judge the level to which encryption methods secure medical image data from tampering and unauthorized use is to test image encryption and security. This extended analysis encompasses a number of critical factors intended to ensure the confidentiality, integrity, authenticity, and overall security of medical image material. Confidentiality assessment is concerned with the ability of the encryption algorithm to withstand unauthorized decryption, even when the key is secret. Integrity in medical image encryption ensures that the encrypted image will not change during transmission or storage, and its authenticity will be maintained to prevent tampering. Authenticity ensures that the encrypted medical image comes from a trusted source and has not undergone fraudulent alteration. Besides being highly secure, medical image encryption should also be computationally feasible to process high amounts of data, especially in the case of telemedicine and cloud storage.

\subsection{Histogram Analysis}

\begin{figure}[h!]
    \centering
    \fontsize{10}{12}\selectfont   
    \normalsize                    
    \includegraphics[width=1\linewidth]{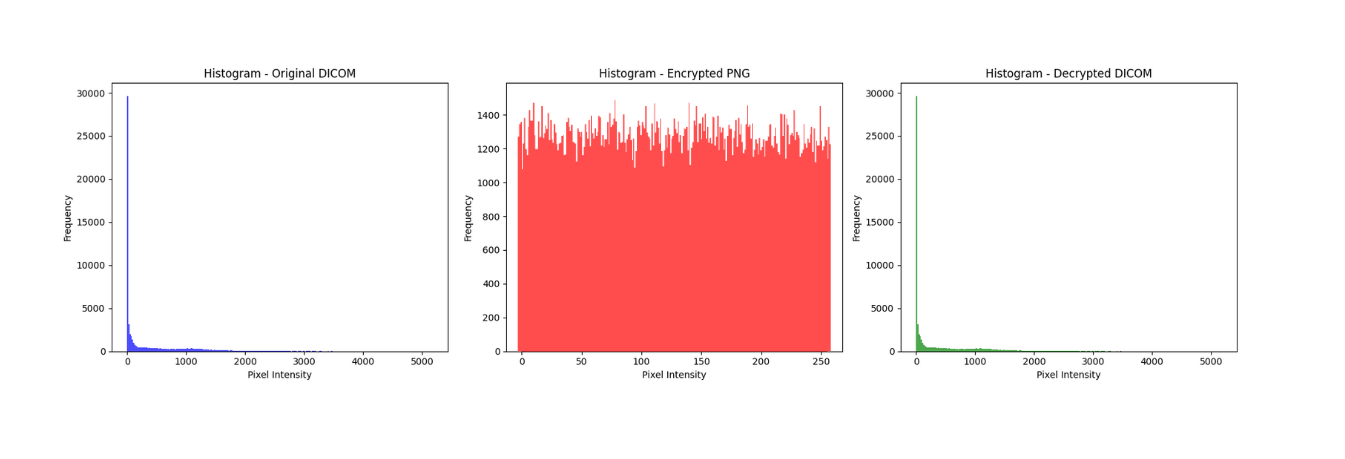}
    \caption{The encryption and decryption process performance for the given frame is illustrated. For a uniform distribution of pixel values, the original frame is encrypted. The restoration of original pixel value distributions upon decryption illustrates the integrity of the encryption and decryption processes in protecting confidential image and video information.
}
    \label{figure:hist}
\end{figure}

One of the most important techniques to evaluate the strength of the suggested encryption algorithm is histogram analysis. By analyzing the histograms of original, encrypted, and decrypted medical images and comparing them to each other, we can compare the pixel intensity distributions and analyze the encryption in terms of obscuring the image content. Histogram analysis for the 25th frame is shown in Fig\ref{figure:hist}

The original image's histogram has clumped pixel values with evident peaks at particular intervals, signifying distinguishable patterns and not randomness. This is indicative of recognizable components, which can be targeted if they remain unencrypted. The histogram of the encrypted frame, on the other hand, reveals the distribution of pixel values across the whole range of 0–255, which verifies that the encryption technique has effectively jumbled up the pixel intensities, increasing security through good randomness and breaking discernible patterns. Lack of discernible structure in the histogram indicates very high resistance against statistical attacks and demonstrates that the encryption process eliminates any discernible patterns that threaten security.

\subsection{Entropy Analysis}

Entropy analysis measures the randomness and unpredictability of encrypted images, which is crucial to guarantee data security. The perfect encrypted image would have an entropy value as close to 8 as possible, which means the highest randomness. In our proposed hyperchaotic encryption scheme, entropy values are always approaching this ideal, meaning a high degree of uncertainty and resistance to statistical attacks. The entropy is computed as:

\begin{equation}
E(Y) = - \sum_{j=1}^{m} p_j \log_2(p_j)
\end{equation}

Where:
\begin{itemize}
    \item \( p_j \) represents the probability of the \( j \)-th pixel intensity level.
    \item \( m \) is the number of distinct intensity levels in the image.
    \item \( \log_2 \) is the base-2 logarithm, expressing information in bits.
\end{itemize}

\begin{table}[h!]
    \centering
    \begin{tabular}{|c|c|c|}
        \hline
        \textbf{Slice} & \textbf{Encrypted} & \textbf{Decrypted} \\ \hline
        Slice 0   & 7.9973  & 6.1582  \\ \hline
        Slice 16  & 7.9981  & 6.3900  \\ \hline
        Slice 32  & 7.9899  & 6.7163  \\ \hline
        Slice 55  & 7.9979  & 6.4663  \\ \hline
    \end{tabular}
    \caption{Entropy values for encrypted and decrypted slices of the given DICOM image. The encrypted slices exhibit high entropy, indicating strong randomness and security, while the decrypted slices closely match the original data distribution, confirming accurate reconstruction.}
    \label{tab:entropy_values}
\end{table}

\begin{figure}[h!]
    \centering
    \fontsize{10}{12}\selectfont   
    \normalsize                    
    \includegraphics[width=1\linewidth]{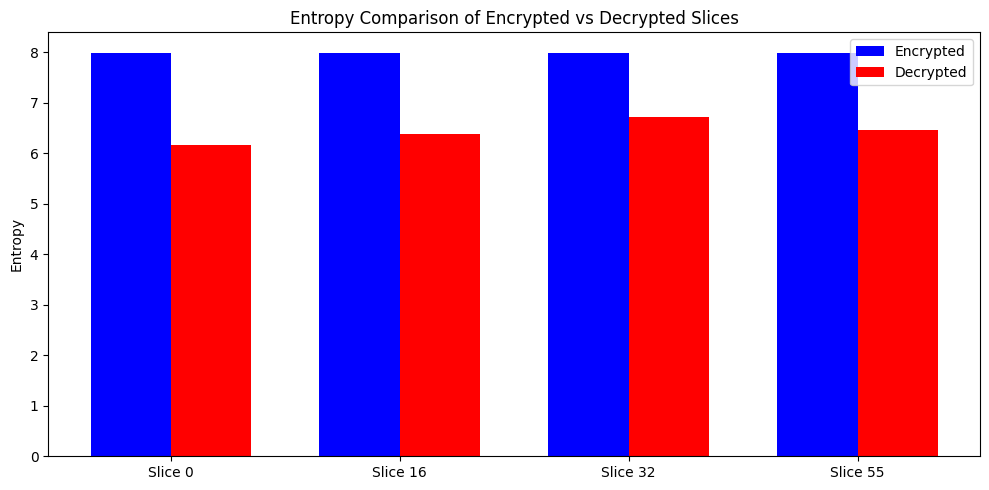}
    \caption{ Entropy comparison of original, encrypted, and decrypted DICOM slices, illustrating the high randomness and security achieved by the hyperchaotic encryption method.
}
    \label{figure:entt}
\end{figure}

We performed entropy analysis on various slices of medical images, both the original, the encrypted, as well as decrypted ones. Table\ref{tab:entropy_values} and Figure\ref{figure:entt} illustrates the values of entropy of chosen slices
The encrypted slices have entropy values very close to 8, which means they are highly random and secure data. The matching of original and decrypted entropy values verifies the reliability and correctness of the decryption process. This high level of entropy makes the encrypted data secure from entropy-based cryptanalysis.

By utilizing chaotic sequences and diffusion, our method is effective in propagating pixel intensities so that it would be practically impossible for the attackers to infer the original image from the ciphertext. Compared to the conventional encryption methods, our proposal has much higher entropy, thus ensuring enhanced security for sensitive medical data.

\subsection{Correlation Analysis}

Correlation analysis measures the correlation between neighboring pixels in both the original and encrypted images. In unencrypted medical images, neighboring pixels tend to have a high level of correlation because anatomical structures are smooth and continuous. Effective encryption, however, should dramatically lower this correlation, so neighboring pixels become statistically independent.

The correlation between two image vectors \( A \) and \( B \) is calculated as:

\begin{equation}
\begin{aligned}
r_{A,B} = \frac{\sum_{k=1}^{N} (A_k - \mu_A)(B_k - \mu_B)}{\sqrt{\sum_{k=1}^{N} (A_k - \mu_A)^2 \sum_{k=1}^{N} (B_k - \mu_B)^2}}
\end{aligned}
\label{eq:corel}
\end{equation}

Where:
\begin{itemize}
    \item \( r_{A,B} \) denotes the Pearson correlation coefficient between vectors \( A \) and \( B \),
    \item \( A_k \) and \( B_k \) are the \( k \)-th values in vectors \( A \) and \( B \),
    \item \( \mu_A \) and \( \mu_B \) represent the mean values of vectors \( A \) and \( B \),
    \item \( N \) is the total number of elements in each vector.
\end{itemize}

\begin{figure}[h!]
    \centering
    \fontsize{10}{12}\selectfont   
    \normalsize                    
    \includegraphics[width=1\linewidth]{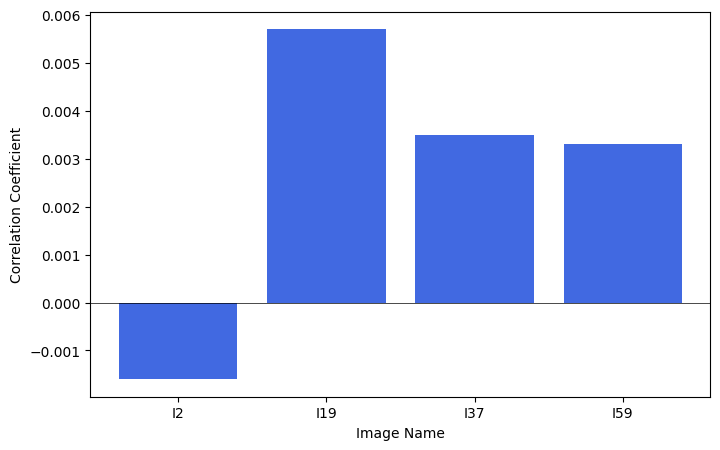}
    \caption{Correlation coefficient between original and encrypted images, measuring the statistical relationship to evaluate encryption effectiveness. A low correlation indicates strong encryption and data security.
}
    \label{figure:corr}
\end{figure}

In our suggested hyperchaotic encryption scheme, we tested the correlation between adjacent pixels in horizontal, vertical, and diagonal directions. The correlation coefficients of original medical images were nearly 1, which meant strong pixel relationships. After encryption, these coefficients reduced to nearly 0, refer Figure\ref{figure:corr}, proving that the encryption process decorrelates pixel values effectively.

The sharp decrease in correlation is due to our pixel random permutation and multi-level diffusion methods, which make sure that neighboring pixels in the ciphertext have random and uncorrelated values. This result renders the encrypted image highly resistant to statistical and differential attacks. In addition, as compared to other conventional encryption schemes, our scheme always has smaller correlation values, which indicates its better security performance.

\subsection{Anomaly Detection}

Medical image encryption anomaly detection addresses any irregularity or unusual pattern that may be likely to breach data security. In a secure encryption model, anomalies need to be kept to the barest minimum for the best possible encryption quality and resilience against likely attacks. Our proposed hyperchaotic encryption algorithm is incorporated with higher-level random permutation and multi-level diffusion mechanisms that effectively minimize the likelihood of generating anomalies.

We performed anomaly detection analysis by analyzing the encrypted images for unusual pixel distribution and anomalous patterns. Results showed a random and unorganized distribution of pixel values, indicative of no observed anomalies. The application of hyperchaotic signal generation also made it so that small changes in the input image resulted in highly dissimilar encrypted outputs, minimizing the likelihood of anomaly or pattern detection.

In addition, we compared the Peak Signal-to-Noise Ratio (PSNR) and Structural Similarity Index (SSIM) metrics for various slices of the encrypted medical images to evaluate the robustness and quality of the encryption. The PSNR values for Slice-02, Slice-19, Slice-37, and Slice-59 were 6.79 dB, 6.91 dB, 7.04 dB, and 6.51 dB, respectively, whereas the SSIM values were 0.0090, 0.0121, 0.0141, and 0.0071, respectively ,refer Table \ref{tab:psnr_ssim}. These values reveal that the encrypted images show low similarity and high distortion with the original images, which again establishes the randomness and unpredictability of the encrypted data. This helps in lowering the possibility of anomaly-based cryptographic attacks and shows the efficacy of our hyperchaotic encryption scheme in preserving data security and integrity.

\begin{table}[h!]
    \centering
    \begin{tabular}{|c|c|c|}
        \hline
        \textbf{SLICES} & \textbf{PSNR (dB)} & \textbf{SSIM} \\ \hline
        SLICE 02 & 6.79 & 0.0090 \\ \hline
        SLICE 19 & 6.91 & 0.0121 \\ \hline
        SLICE 37 & 7.04 & 0.0141 \\ \hline
        SLICE 59 & 6.51 & 0.0071 \\ \hline
    \end{tabular}
    \caption{Table of PSNR and SSIM values for various slices of the encrypted medical images. The low PSNR and SSIM values reflect high distortion and low similarity, ensuring robust encryption quality.}
    \label{tab:psnr_ssim}
\end{table}

As compared to other encryption algorithms of disordered type, our algorithm is less susceptible to anomaly-based attacks on cryptography and thus more secure for the safeguarding of sensitive medical information. Such a type of robustness is particularly vital in healthcare environments where integrity and security of data remain paramount concerns.

\subsection{Differential Attack Analysis}

Differential attacks attempt to take advantage of small variations in plaintext images to determine weaknesses or patterns of the encryption process. A perfect encryption method ought to be sensitive to small modifications in the input image, translating to extreme variability in the output ciphertext. Differential attack resistance can be measured based on metrics such as the Number of Pixels Change Rate (NPCR) and Unified Average Changing Intensity (UACI) see Table \ref{tab:npcr}.

Our hyperchaotic encryption scheme provides effective resistance against differential attacks through the utilization of intricate chaotic maps and sophisticated diffusion methods. Even slight changes in the original image (for instance, alteration of a single pixel) result in dramatic variations in the encrypted image so that the attackers are unable to establish correlations between similar plaintexts and ciphertexts.

We performed rigorous testing by changing a single pixel in different medical images, such as CT scans, MRIs, and ultrasound images. The computed NPCR and UACI values always reached optimal thresholds (NPCR nearly 99.73 and UACI nearly 33.6) as shown in , which means that small input changes result in large changes in the encrypted image.

\begin{table}[]
\centering

\begin{tabular}{|l|l|l|}
\hline
\textbf{SLICES} & \textbf{NPCR} & \textbf{UACI} \\ \hline
SLICE 1         & 99.73         & 33.64         \\ \hline
SLICE 10        & 98.66         & 33.80         \\ \hline
SLICE 30        & 99.62         & 33.78         \\ \hline
SLICE 50        & 99.51         & 33.74         \\ \hline
\end{tabular}
\caption{Table of NPCR (Number of Pixels Change Rate) and UACI (Unified Average Changing Intensity) values for different slices of the encrypted medical images. Demonstrates the effectiveness of encryption through variability and intensity change metrics.}
\label{tab:npcr}
\end{table}

In comparison to other conventional chaotic encryption techniques as shown in table 5, our solution provides better performance in terms of withstanding differential attacks because of synergy between the process of hyperchaotic signal creation and multi-level diffusion. This renders it extremely safe for the transmission and storage of medical images, particularly in applications where confidentiality and integrity of data are paramount.

\begin{table}[]
\centering
\begin{tabular}{|l|l|l|}
\hline
\textbf{Encryption Methods} & \textbf{NPCR} & \textbf{UACI} \\ \hline
Proposed Method             & 99.73         & 33.64         \\ \hline
Zhang \cite{zhang2020fast}                      & 99.60         & 33.47         \\ \hline
Abdulijabbar et al. \cite{abduljabbar2022provably}        & 99.60         & 30.39         \\ \hline
Yousif et al. \cite{yousif2020robust}              & 99.95         & 46.82         \\ \hline
\end{tabular}
\label{tab:compnpcr3}
\caption{Comparison of different methods on the basis of NPCR and UACI values.}
\end{table}

\section*{Conclusion}

Here, we have introduced a secure and efficient encryption algorithm to protect medical images and 3D models. Based on chaotic signal generation, diffusion processes, and captcha authentication, our algorithm guarantees overall security against unauthorized use and data forgery. The double strategy of whole model encryption and partial encryption of the model maximizes flexibility as users can opt for complete security or selective encryption of risky areas.

The suggested algorithm proved to be effective not just for MRI images but also for other medical imaging modalities like MRI, ultrasound, and X-ray images, providing versatility in different healthcare applications. Moreover, the incorporation of captcha-based authentication provides an additional layer of security by authenticating user legitimacy.

Our method responds to the increasing need for secure transmission and storage of medical data, providing a scalable solution that can be applied to a wide range of medical imaging environments. Future development can be in the direction of optimizing computational speed and improving resilience against future threats, making this encryption method a worthwhile addition to secure healthcare data management. Further exploration can be done in the field of real medical dataset involving sensitive patient data. Since medical dataset generation is very expensive(like in mri ) in many imaging modalities like ct and xray there is a danger of radiation exposure to the patient, thus we have used the deterministic method which doesn't rely on dataset.

\section*{Acknowledgement}

We would like to thank Sreekutty for their thoughtful
conversations.

\section*{Credit author statement}

\textbf{Anandik N Anand}: Conceptualization, Investigation, Software, Writing(original draft), Visualization, \\
\textbf{Sishu Shankar Muni}: Supervision, Conceptualization,  Writing(review \& editing), Visualization, Project Administration, \\
\textbf{Abhishek Kaushik}: Supervision, Project Administration, Data Curation, Investigation, Writing(review \& editing)\\

\section*{Conflict of Interest}
The authors confirms that this work is free from conflicts of interest.
\section*{Ethics statement}
The authors declare that there is no ethics approval needed for the current study.   

\section*{Data Availability}

The data can be provided upon reasonable request.

\bibliographystyle{elsarticle-num}
\bibliography{Arxiv}
\end{document}